**The apparent fine-tuning of the cosmological, gravitational and fine structure constants**

Laurence Eaves

*School of Physics & Astronomy, University of Nottingham, Nottingham NG7 2RD, UK*

**Abstract**

A numerical coincidence relating the values of the cosmological, gravitational and electromagnetic fine structure constants is presented and discussed in relation to the apparent anthropic fine-tuning of these three fundamental constants of nature.

**1.    Introduction**

Understanding the physics of the cosmological constant, $\Lambda$, remains one of the outstanding challenges in science.  If $\Lambda$ were significantly smaller than its measured value [1-3], it would be too small to detect with present technology, whereas its upper bound is constrained by arguments based on the anthropic principle and cosmology [4-11].  Using a set of four statistical axioms, Beck [12] has argued that the value of $\Lambda$ is determined by gravitational and electromagnetic interactions rather than short-range forces.  He thereby obtained a formula for $\Lambda$ in terms of the low energy value of the electromagnetic fine structure constant, $\alpha$, the gravitational constant, $G$, Planck's constant $\hbar$, the electronic mass, $m_e$, and charge, $e$.  Here we consider Beck's result and a numerical coincidence that inter-relates the measured values of Λ, G and α. Together they suggest that the apparent fine-tuning of $\alpha$ would also ensure that Λ and G have the values that we measure for our universe.

2.  Discussion

Beck's formula for Λ can be written in the following form (the subscript B distinguishes it from the measured value of Λ):

$$2\pi t_e \Lambda_B^{1/2} = \left(\frac{L_P}{r_e}\right)^2. \tag{1}$$

Here $L_P = \sqrt{hG/c^3} = 4.05 \times 10^{-35}$m is the Planck length, $r_e = e^2/4\pi\varepsilon_0 m_e c^2 = 2.82 \times 10^{-15}$ m is the classical electron radius, $t_e = r_e/c$ is the corresponding time. Beck's value of Λ, which agrees with that obtained from astrophysical data to within experimental uncertainty, involves the ratio $N = (r_e/L_P)^2 \sim 10^{40}$. This large number determines the relative strengths of the electrical and gravitational forces between two electrons and is associated with the famous discussion involving Dirac, Dicke and others concerning the approximate coincidence between the value of N and the present age of the universe, $T_U$, in units of $t_e$ [13-15]. Dicke explained the coincidence by invoking the anthropic principle: a universe that can support observers inhabiting rocky planets must be at least as old as the lifetime, $t_s$, of a typical solar mass star. According to astrophysical arguments, $t_s \simeq t_e N K$, where $K = 4\pi m_e/3\alpha m_N$ is a numerical constant of order unity and $m_N$ is the nucleon mass [6,16,17]. Hence Beck's $\Lambda_B$ is consistent with the relation $\Lambda_B^{1/2} t_s \sim 1$ and with the so-called "coincidence problem", $\Lambda_B^{1/2} T_U \gtrsim 1$. Another cosmological coincidence relates Beck's expression for Λ to G and $r_e$: a small black hole with Schwartzchild radius $r_e$ and mass $M_{BH}^e = r_e c^2/2G \sim 10^{12}$kg has a Hawking radiation-limited lifetime $\tau_{BH}^e \sim N t_e$ (see relation 13 of ref. 6), i.e. black holes of radius $r_e$ that formed in the early universe are ending their lives in the present epoch when $T_U \sim \Lambda^{-1/2} \sim \tau_{BH}^e$.

Whereas Beck's result indicates a constant value for $\Lambda$, Barrow and Shaw have recently used the gravitational action principle to deduce that classical observers would measure $\Lambda$ to have a value such that $\Lambda^{-1/2}$ is close to the age of the universe at the time of their observations [18,19], thus providing an explanation of the coincidence problem. Sorkin [20] obtained a similar result using the uncertainty principle and a statistical argument: on the Planckian length scale $L_P$, the local value of $\Lambda$ can undergo random fluctuations of magnitude $\Delta\Lambda \sim (c/L_P)^2$. Since our observable universe contains $N_2 \approx (cT_U/L_P)^4 \sim 10^{240}$ Planckian boxes, the typical value of $\Lambda$ measured on a cosmological scale would then average out to be of magnitude $\Delta\Lambda/\sqrt{N_2}$, so that $\Lambda \sim T_U^{-2}$.

Since Beck's $\Lambda_B$ involves $\alpha$ and $G$, let us recall earlier discussions of how these constants appear to be related to each other. The logarithmic relation, $\alpha^{-1} \sim \log \alpha_G^{-1}$, where $\alpha_G = Gm_N^2/\hbar c$ is the conventionally defined form of the gravitational fine structure constant, has been long regarded as a requirement for a self-consistent electrodynamics [6,21-24]. Barrow and Tipler have also discussed approximate relations involving $e^{1/\alpha}$ and cosmological parameters, see chapter 5 of ref. [7]. A more recent renormalisation group analysis by Page of supersymmetric grand unified theories suggests that $\alpha^{-1} \approx (5/\pi)\ell n\, \alpha_G^{-1}$ [25]. We now introduce a hypothetical yet accurate numerical coincidence which involves $N = (r_e/L_P)^2$. It suggests the following relation between $G$, $\alpha$, $m_e$ and $q^2 = e^2/4\pi\varepsilon_0$:

$$N = \left(\frac{r_e}{L_P}\right)^2 = \frac{\alpha q^2}{2\pi G m_e^2} \approx e^{2/3\alpha}. \tag{2}$$

The accuracy of this approximation can be seen by noting that it becomes an equality if $\alpha^{-1}$ is adjusted to be 137.066, which is larger than its measured value of 137.036 (rounded to 6 significant figures), by only ~0.02% (~$\alpha^2$). The relation $N \approx e^{2/3\alpha}$ can also be written in integral form, $\int_{V_P}^{V_e} dV/V \approx \alpha^{-1}$, in which the upper and lower limits are the classical electron and Planck volumes.

By combining (2) with Beck's relation (1), we can eliminate $G$ and write $\Lambda_B$ in terms of electromagnetic parameters only:

$$2\pi t_e \Lambda_B^{1/2} \approx e^{-2/3\alpha}. \tag{3}$$

Relations (2) and (3) take the following simple forms when expressed in electron units (e.u.), with $r_e = m_e = q^2 = c = 1$ and $\alpha^{-1} = \hbar$:

$$(2\pi G)^{-1} \approx \hbar e^{2\hbar/3} \tag{4}$$

and

$$2\pi \Lambda_B^{1/2} \approx 1/\tau_{BH}^e \approx e^{-\frac{2\hbar}{3}}. \tag{5}$$

The corresponding expressions for the Planck energy $\varepsilon_P \approx \hbar e^{\hbar/3}$ and the energy $\varepsilon_{\Lambda_B} = h\Lambda_B^{1/2} \approx \hbar e^{-2\hbar/3}$ both have the form of the Lambert function, $W(z)$, given by $z = We^W$.

The apparent fine-tuning of the fundamental constants have been discussed intensively in relation to the so-called anthropic principle [5-11, 23-26] and the multiverse scenario [27-29]. In particular, a value of $\alpha^{-1}$ close to 137 appears to be essential for the astrophysics, chemistry and biochemistry of our universe. With such a value of $\alpha$, the exponential forms of relations (2) and (3) indicate that $\Lambda$ and $G$ have the extremely small values that we observe in our universe, so that $\Lambda$ is small enough to permit the formation of large-scale structure, yet large enough to detect and measure with present-day astronomical technology, and $G$ is small enough to provide stellar lifetimes that are sufficiently long for complex life to evolve on an orbiting planet, yet large enough to ensure the formation of stars and galaxies. Thus, if Beck's result is physically valid and if the hypothetical relation (2) represents an as yet ill-defined physical law relating $\alpha$ to $G$, then the apparent fine-tuning of $\Lambda, G$ and $\alpha$ is reduced to just one, rather than three coincidences, of seemingly extreme improbability.

In a multiverse scenario [27], one can envisage a subset of universes containing electrons, but in which $\Lambda$, $G$ and $\hbar$ take on different values from those in our universe. In that case, electron units seem more appropriate than Planck units or the classical units proposed in 1881 by Johnson Stoney and based on $G$, $c$ and $e$ [30,31]: one can then speculate about the implications of relations (4) and (5) for hypothetical universes which contain electrons but in which $\alpha^{-1}$ differs significantly from 137. Such universes would have exponentially different values of $G$ and $\Lambda$ and be inimical to life as we know it.

3. **Conclusions**

There are well-reasoned arguments that the multiverse concept [29] is unscientific. Similar criticisms have been made of the anthropic principle and the use of numerical coincidences. However, equations (1) and (2) inter-relate, albeit empirically, a set of three precisely-measured fundamental constants of nature to a surprising degree of accuracy. The history of science provides a reminder of the usefulness of empirical hypotheses: for example, a numerical law for the wavelengths of the spectral lines of hydrogen was developed long before the formulation of quantum mechanics [32,33].